\documentclass{article}[12pt]
\input{epsf}
\begin{document}
\newcommand{\And}{\textup{and}\ \ignorespaces}
\let\atque\And
\newcommand{\drm}{\mathrm{d}}        %MATH ONLY
\newcommand{\sy}[1]{\mathrm{#1}}     %MATH ONLY
\let\ab\sy
\newcommand{\tx}[1]{\mbox{#1}}
\newcommand{\un}[1]{{\,\mbox{#1}}}
\newcommand{\chem}[1]{\ensuremath{\mathrm{#1}}}
\newcommand{\acro}[1]{{\normalfont #1}}
\newcommand{\vcomment}[1]{%
  \unskip
  \ifcomments
    \marginpar{%
      \hrule height2pt
      \vspace*{4pt}
      {\raggedright\footnotesize\itshape #1\par}%
      \vspace*{4pt}
      \hrule height2pt
    }%
  \fi
}
\def\@deceased@{\let\@temp@\@makefnmark
                \def\@makefnmark{$^\dag$}\thanks{Deceased.}%
        \let\@makefnmark\@temp@}
\newcommand{\deceased}{\protect\@deceased@}
%%%%%%%%%%%%%%%%%%%%%%%%%%%%%%%%%%%%%%%%%%%%%%%%%%%%%%%%%%
%% It's not a dumb class anymore ;( (isn't it?)
 \newcommand{\tocex}[3]{%
     \textsc{##2}~--~##3~\textit{(\@@xtr@@)}\dotfill%
       \ifnum\value{pageflag}=0
       \rlap{\hbox to1cm{\quad\hfill page\hfill}\hbox to1cm{\hfill##1}}%
     \else
       \rlap{\hbox to1cm{\quad\hfill$\scriptstyle{``}$\hfill}\hbox to1cm{\hfill##1}}%
     \fi
     \stepcounter{pageflag}
     \par
     \pagebreak[3]%
   }%
   \newcommand{\tocit}[3]{%
     \textsc{##2}~--~##3\dotfill%
       \ifnum\value{pageflag}=0
       \rlap{\hbox to1cm{\quad\hfill page\hfill}\hbox to1cm{\hfill##1}}%
     \else
       \rlap{\hbox to1cm{\quad\hfill$\scriptstyle{``}$\hfill}\hbox to1cm{\hfill##1}}%
     \fi
     \stepcounter{pageflag}
     \par
     \pagebreak[3]%
   }%
   \newcommand{\tocem}[2]{%
     \tocskip
     {##2\dotfill}%
       \ifnum\value{pageflag}=0
       \rlap{\hbox to1cm{\quad\hfill page\hfill}\hbox to1cm{\hfill##1}}%
     \else
       \rlap{\hbox to1cm{\quad\hfill$\scriptstyle{``}$\hfill}\hbox to1cm{\hfill##1}}%
     \fi
     \stepcounter{pageflag}
     \par
     \pagebreak[3]%
   }%
   \newcommand{\tocsect}[2]{%
     \tocskip
     \tocskip
     ##2
     \hskip4pt##1
     \tocskip
     \par
   }%
   \newcommand{\tocskip}{\medskip}%
   \newcommand{\shortnotes}{%
     \tocskip
     \textbf{Note brevi}%
     \par
   }%
%%%%%%%%%%%%%%%%%%%%%%%%%%%%%%%%%%%%%%%%%%%%%%%%%%%%%%%%
% The title, all uppercase; if you want to split it in
% two or more lines, put a \\ macro at each line break
% example:
%   \title{TITLE: FIRST LINE\\ SECOND LINE}
%
\title{Physical restrictions on the models of gamma-ray bursts}
%%%%%%%%%%%%%%%%%%%%%%%%%%%%%%%%%%%%%%%%%%%%%%%%%%%%%%%%
% The author(s), separated by commas; do not put a
% comma before the last author, use instead the \And
% macro which produces a normal ``and'' in the
% caps/small caps context
%
\author{Guennadi Bisnovatyi-Kogan\thanks{Space Research Institute Rus. Acad. Sci., Moscow,
Russia}}
%%%%%%%%%%%%%%%%%%%%%%%%%%%%%%%%%%%%%%%%%%%%%%%%%%%%%%%%
%
\date{}
\maketitle
%%%%%%%%%%%%%%%%%%%%%%%%%%%%%%%%%%%%%%%%%%%%%%%%%%%%%%%%
% Write the text starting from here and using the usual
% LaTeX commands.
%

\begin{abstract}
The present common view about GRB origin is related to cosmology.
There are two evidences in favor of this interpretation. The first is
connected with statistics, the second is based on measurements of the
redshifts in the GRB optical afterglows. Redshifts in optical afterglows
had been observed only in long GRB. Comparison of KONUS and BATSE
data about statistics and hard X-ray lines is done, and some
differences are noted. Hard gamma-ray afterglows, prompt optical
spectra and polarization measurements could be very important for
farther insight into GRB origin.
 Possibility of galactic
origin of short GRB is discussed as well as their possible connection with soft
gamma repeaters.
\end{abstract}

\section{Introduction}

It is generally accepted now that cosmic gamma-ray bursts (GRB) discovered
in 1973 \cite{kle}
have a cosmological origin. The first
cosmological model, based on explosions in active galactic nuclei
(AGN) was suggested in \cite{pu}.
A mechanism of the
GRB origin in the vicinity of a collapsing object based on
neutrino-antineutrino annihilation was analyzed in \cite{bp}.
It was obtained that the efficiency of
transformation of the neutrino flux energy $W_{\nu} \sim 6 \cdot
10^{53}$ ergs into $X$-ray and $\gamma$-ray burst is $\alpha \sim
6 \cdot 10^{-6}$, with the energy output in the GRB $W_{X,\gamma}
\sim 3 \cdot 10^{48}$ ergs. Numerical
three-dimensional simulations of models of two colliding
neutron stars (NS) \cite{rj98}  and of a hot
torus around a black hole (BH) \cite{rj99} gave larger
efficiency of  $X$-ray and $\gamma$-ray production, up to
$0.5 \%$ in the first, and $1 \%$ in the second model. This
difference may be partly connected with more preferable geometry
of the neutrino flux increasing the annihilation rate relative to
estimations in the spherical geometry in \cite{bp}.
 Nevertheless, even these optimistic results, permitting
GRB formation with a total $X$-ray and $\gamma$-ray energy up to
$5 \cdot 10^{50}$ ergs are not enough for explanation of the
energy output in some GRB, where only prompt optical energy emission
reaches $10^{51}$ ergs, and the isotropic gamma-ray flux is about
$2.3 \cdot 10^{54}$ ergs: GRB 990123 with the red shift $z \sim
1.6$ \cite{ake,ku}.

Here we discuss different observational features of GRB,
analyze difficulties and problems of their interpretation in the cosmological
model, and physical restrictions to their model. At the end we are analyze
some problems of soft gamma repeater (SGR) interpretation as magnetars.

\section{GRB physical models}

The restrictions to the model follow from the energy conservation
law, but much stronger ones are imposed by the necessity to
fulfill the physical laws: the weakly interacting neutrino can
be transformed into the radiation only with a rather low efficiency.
The GRB models may be classified by two levels. The upper one is
related directly to the observational appearance, and include 3
main models.

1. Fireball.

2. Cannon ball (or gun bullet).

3. Precessing  jets.

\noindent The main restrictions are connected with the next
(basic) level of GRB model, which is related to energy source,
producing a huge energy output necessary for a cosmological GRB
model. These class
 contains 5 main models.

1. (NS+NS), (NS+BH) mergers.

\noindent
This mechanism was investigated numerically in \cite{rj98, rj99},
Gamma radiation is produced here by $(\nu, \tilde{\nu})$ annihilation,
and the energy output is not enough to explain most powerful GRB
even with account of strong beaming. The energy emitted in the optical
afterglow of GRB 990123 \cite{ake,ku} is about an order of magnitude
larger than the total radiation energy output in this model.

2. Magnetorotational explosion.

\noindent Magnetorotational explosion, proposed in \cite{pa} for
an explanation of the huge energy production in a cosmological
GRB, had been suggested earlier for the supernova explosion in
\cite{bk}. Numerical 1-D and 2-D calculations gave the efficiency
of a transformation of the rotational energy into the kinetic one
at the level of few percent \cite{ard,abm}. This is enough for an
explanation of the supernovae energy output but is too low for
cosmological GRB, because the energy lost by radiation is even
less than in the merger model.

3. Hypernova.

\noindent
This model, also suggested in \cite{pa} is rather popular now,
because traces of the supernova explosions are believed to be
found in the optical afterglows of several GRB \cite{sok,ddd}.
 The restrictions of the
"hypernova" model  had been analyzed in \cite{bl2}.

The first two models seems to be able to produce the energy output
on the level of the ordinary supernova explosion, but there are no
numerical simulations of these explosions, in which the energy
output was enough for a cosmological GRB. The third model is more
vaguely formulated and no numerical simulations have been done so
far.

4. Magnetized disks around rotating (Kerr) black holes (RBH).

\noindent This model is based on extraction of rotating energy of
RBH when magnetic field is connecting the RBH with the surrounding
accretion disk or accretion torus \cite{rw,bz,vpu}.

5. GRB created by the pair-electromagnetic pulse from an
electromagnetic black hole surrounded by a baryonic remnant. This
model \cite{rswx} is based on vacuum explosion in the dyadosphere,
the region in which a supercritical field exists for the creation
of e$^+$e$^-$ pairs. The problem of formation of such region needs
farther clarification.

\section{Basement of a cosmological GRB origin:\\ statistics}

The conclusion about the cosmological origin of GRB is based on
the analysis of their statistical properties, and spectra of
optical afterglows, showing highly redshifted lines.

\begin{figure}
\epsfsize=05.3cm
\epsfbox{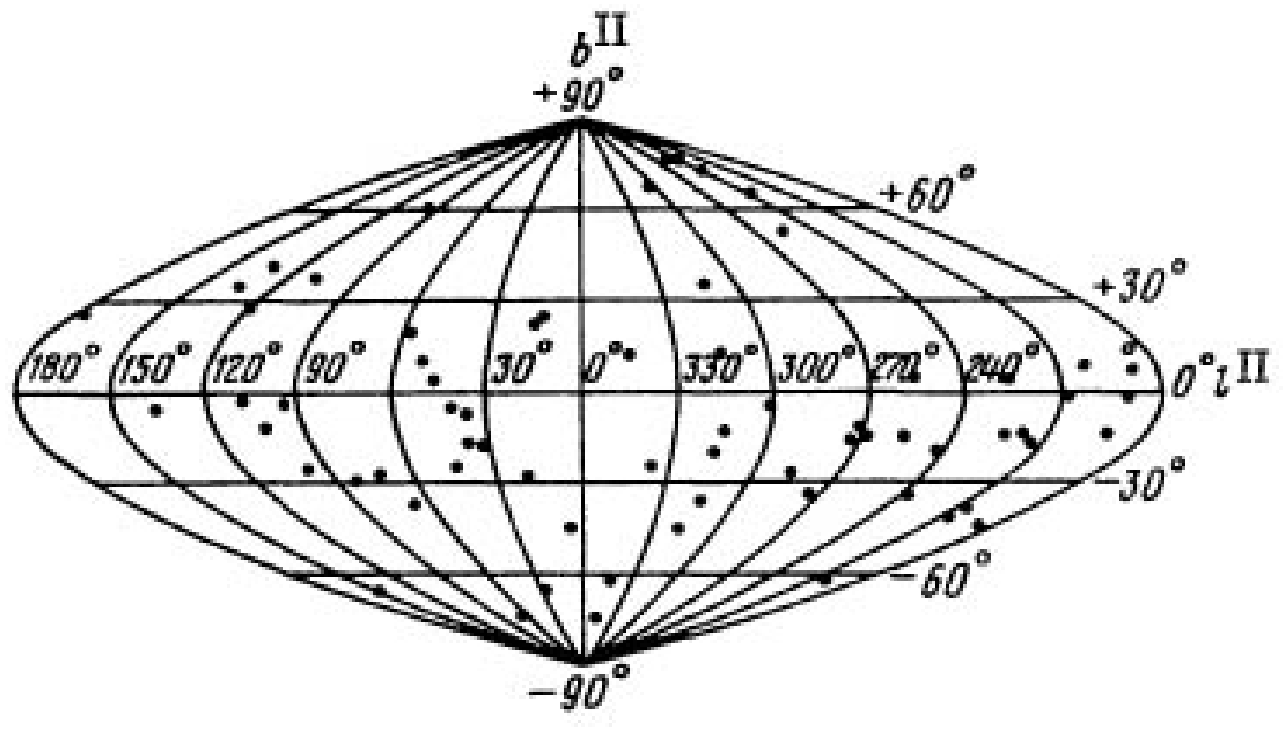}\,\epsfxsize=5.3cm
\epsfbox{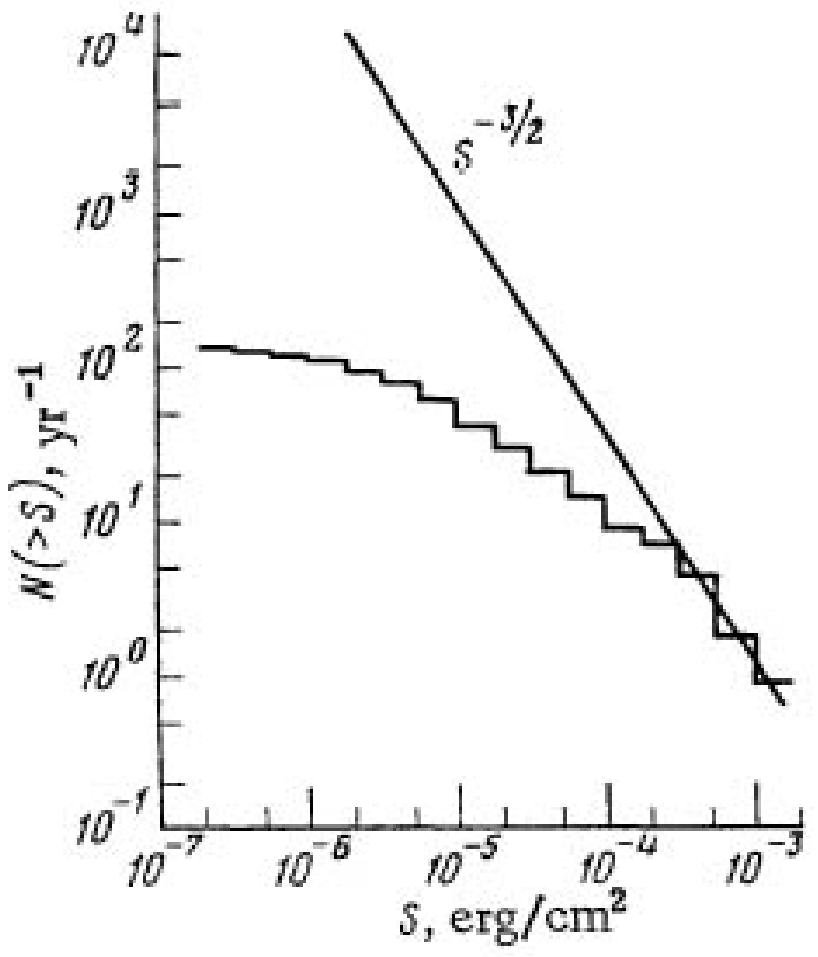}\,\epsfxsize=5.3cm
%{\centerline{\epsfbox{grbfig1.eps}}}
%\vspace{12cm}
\caption[h]{{\bf a.} The position of the GRB sources on the celestial sphere, mapped in
galactic coordinates $l^{II},\, b^{II}$.
{\bf b.} The  $\log N\, -\, \log S$ distribution for the GRB
recorded in the KONUS experiment, from
\cite{kon80}.}
\label{grbfig1}
\end{figure}

Statistical arguments in favor of the cosmological origin of GRB
are based on a visual isotropy of GRB distribution on the sky in
combination with a strong deviation of $\log N\, -\, \log S$ (or
equivalent) distribution from the euclidian uniform distribution
with the slope $3/2$, obtained in BATSE observations \cite{mee}.
This observational result is not new. Similar properties have been
obtained in KONUS experiment \cite{kon80} represented in
fig.{\ref{grbfig1}}.

\begin{figure}[h]
\epsfsize=07cm
{\centerline{\epsfbox{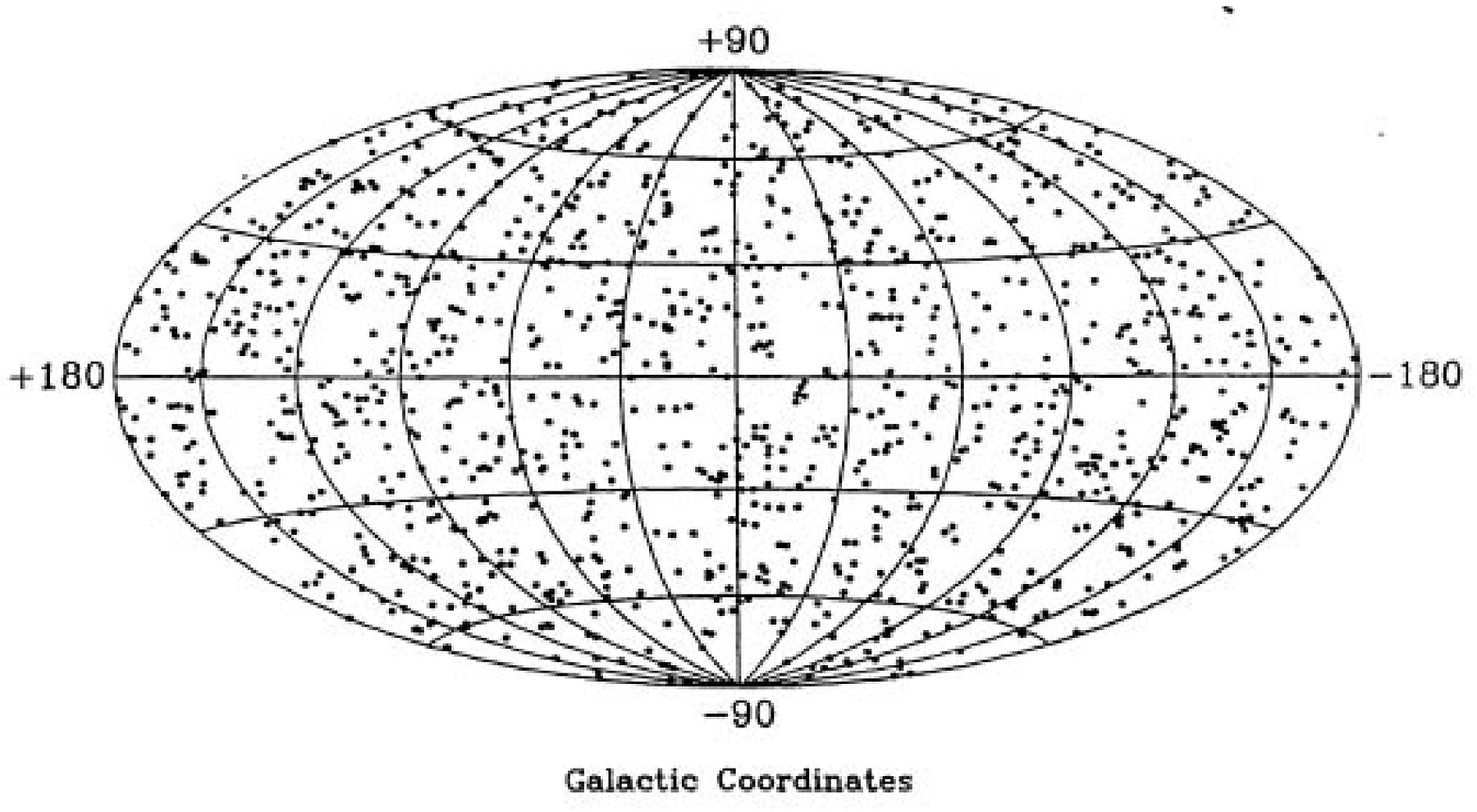}}}
%\vspace{12cm}
\caption[h]{The celestial distribution of 1121 GRB as seen by
BATSE over the three-year period, plotted in galactic coordinates.
No clustering or anisotropy are seen, from \cite{fm95}.}
\label{grbfig2}
\end{figure}

\begin{figure}[h]
\epsfsize=07cm
{\centerline{\epsfbox{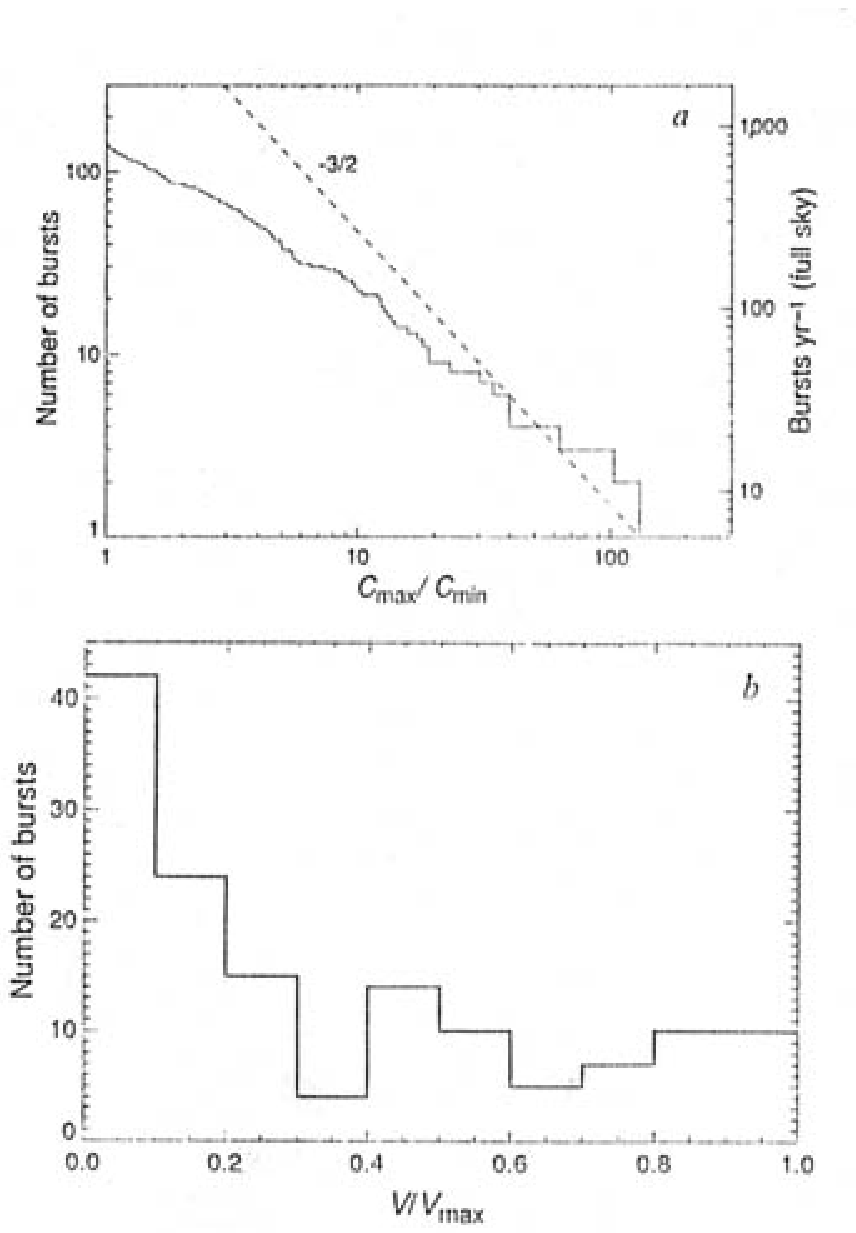}}}
%\vspace{12cm}
\caption[h]{{\bf a}. Integral number distribution of 140 GRB as a
function of peak rate . A -3/2 power law is expected for a
homogeneous distribution of sources. The full sky rate is $\sim
800$ bursts per year. {\bf b}. $V/V_{max}$
distribution for 140 GRB. The average $V/V_{max}$ is
0.348$\pm$0.024. A uniform distribution is expected for a
homogeneous distribution of sources, from \cite{mee}.}
\label{grbfig3}
\end{figure}

\begin{figure}[h]
\epsfsize=06cm
{\centerline{\epsfbox{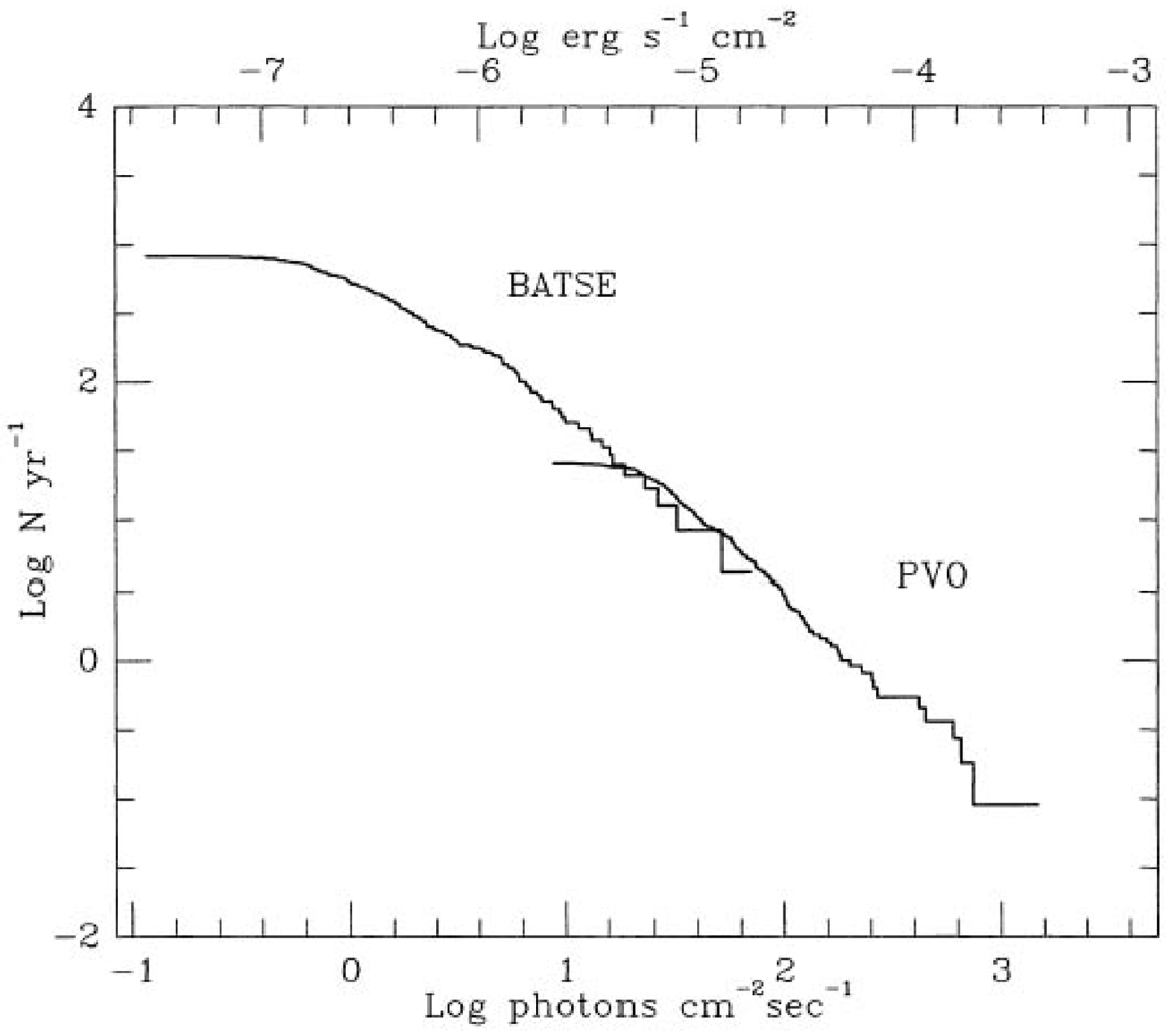}}}
%\vspace{12cm}
\caption[h]{The  $\log N\, -\, \log P$ distribution from combined
BATSE and {\it PVO} data. The distribution match well in the overlap
region. The {\it PVO} data, which has recorded more strong bursts
than BATSE during its long lifetime, is seen to follow a -3/2
power law for strong bursts, from \cite{fm95}.}
\label{grbfig4}
\end{figure}

The authors suggested that
properties of $\log N\, -\, \log S$ curve are connected with
different selection effects, and the actual density of GRB is
almost uniform in space. The account of selection effects in KONUS
experiment made in \cite{hs90} gave the average value  $<V/V_{max}>=0.45 \pm
0.03$; the value 0.5 corresponds to the pure uniform distribution.
KONUS data had been obtained in conditions of constant background.
Similar analysis \cite{s99} of BATSE data, obtained in conditions
of substantially variable background, gave resulting
$<V/V_{max}>=0.334 \pm 0.008$. The statistical
analysis of BATSE data  \cite{mee,fm95} is represented in
figs.\ref{grbfig2},\ref{grbfig3}.
These two results seems
to be in contradiction, because KONUS sensitivity was only 3 times
less than that of BATSE, where deviations from the uniform
distribution $<V/V_{max}>=0.5$ in BATSE data are still rather large
\cite{fm95}. The results of combined BATSE - PVO data are
represented in fig.\ref{grbfig4} from \cite{fm95}. PVO data where
the most luminous GRB are present show good uniformity with a nice
slope of $3/2$.

Detailed statistical analysis and calculation of
of BATSE data, divided in 4 classes
according to their hardness and calculation of  $<V/V_{max}>$ for
different classes
have been done by M. Schmidt \cite{s01}.
The results of this investigation are shown in Table 1, where
$\alpha_{23}$ determines the photon spectrum slope, derived from
counts of BATSE channels 3 (100-300 keV) and 2 (50-100 keV).
The columns "obs" represent observational data corrected for the
effects of statistical errors in the peak counts, and in the "corr"
columns the data had been reanalyzed under suggestion of existence
of luminosity-hardness correlation.

\begin{table}
\caption{Dependence of ${<V/V_{max}>}$ on hardness ratio for 1391 GRBs, from \cite{s01}}
\begin{center}
\medskip
\begin{tabular}{ccccc}
\hline\noalign{\smallskip}
{number}&
{$<\alpha_{23}>_{obs}$}&
{${<V/V_{max}>}_{obs}$}&
{$<\alpha_{23}>_{corr}$}&
{${<V/V_{max}>}_{corr}$}\\
\noalign{\smallskip}
\hline\\
 348 & $-2.55$ & $0.468 \pm 0.017$ & $-2.33$ & $0.421$ \\
 348 & $-1.84$ & $0.309 \pm 0.016$ & $-1.79$ & $0.325$ \\
 347 & $-1.47$ & $0.299 \pm 0.016$ & $-1.47$ & $0.344$ \\
 348 & $-1.04$ & $0.270 \pm 0.015$ & $-1.10$ & $0.256$ \\
\hline\\
\end{tabular}
\end{center}
\end{table}

In the cosmological model we may expect smaller value of
$<V/V_{max}>$ for softer GRB in the case of a uniform sample, because
larger red shifts would correspond to softer spectra. The result
is quite opposite, and soft GRB have larger  $<V/V_{max}>$ than the
hard ones, 0.47 and 0.27 respectively.

\noindent It is supposed in \cite{s01}
such a strong excess of luminosity in hard GRB, which overcomes
the tendency of the uniform sample. Another explanation in which
the soft GRB sample is more complete than the hard one seems to me
more preferable.

\begin{figure}[h]
\epsfsize=06cm
{\centerline{\epsfbox{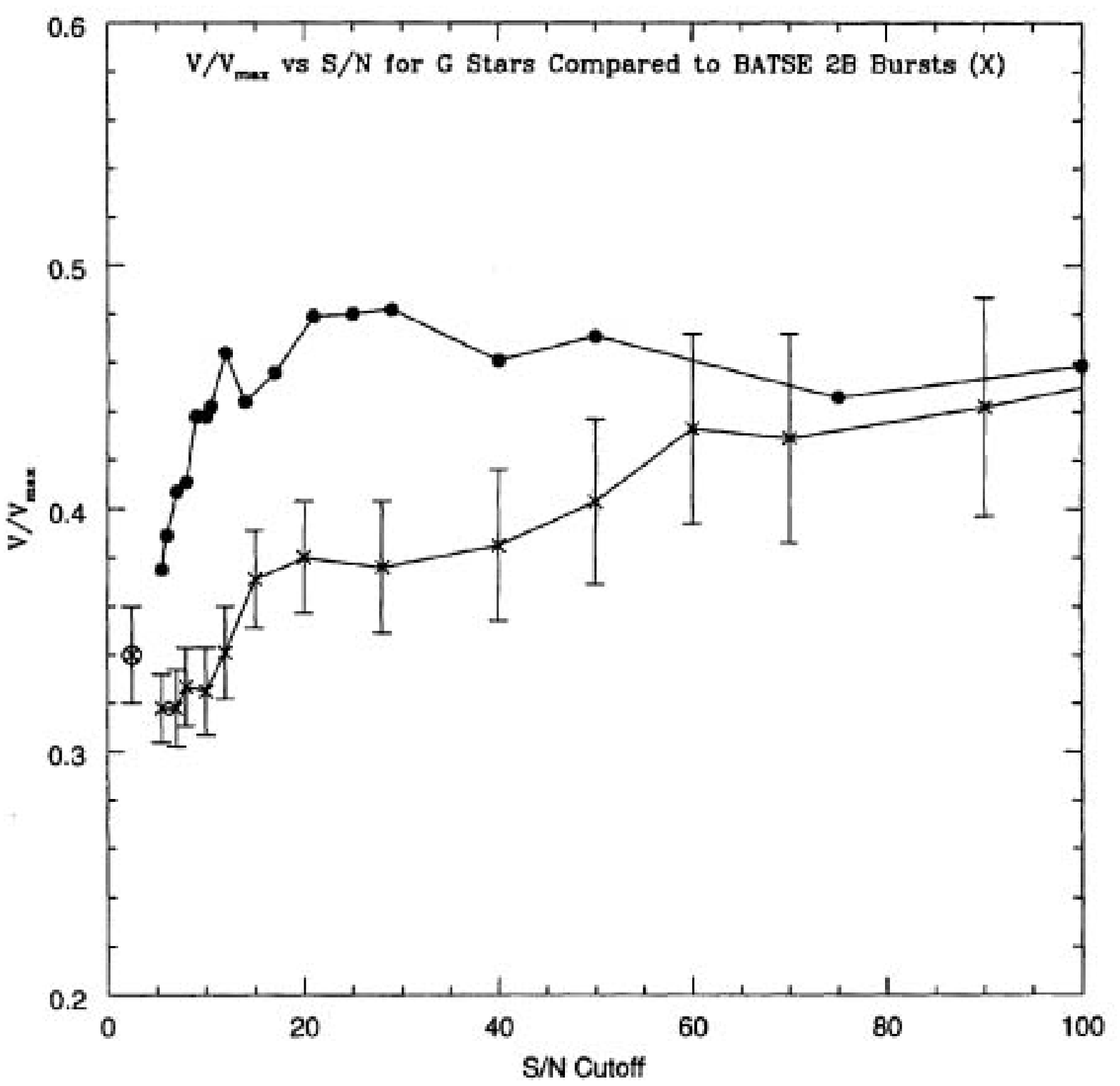}}}
%\vspace{12cm}
\caption[h]{The distribution by brightness of the $V/V_{max}$
parameter for G stars (solid circles), and GRBs from BATSE 2B
catalogue (crosses with 1$\sigma$ error bars). The error bars for
G stars are smaller than the size of the symbols. The circled
cross is the value of $V/V_{max}$ from the BATSE 1B catalogue.
BATSE imposed a 5.5$\sigma$ limit for burst detection. The set
$m_v=10$ for G stars equal to $S/N=5.5$ was used in the
present analysis. Note that for both bright bursts and G stars
$V/V_{max}$ is close to the value for homogeneity (i.e.,
$<V/V_{max}>$=0.50). As fainter sources are sampled, however
$V/V_{max}$ deviates to lower values. For G stars this is due to
incompleteness of the catalogue at faint magnitudes, from \cite{har95}.
} \label{grbfig5}
\end{figure}

\begin{figure}[h]
\epsfsize=06.5cm
%{\centerline{\epsfbox{grbfig7.eps},grbfig8.eps}}}
\epsfbox{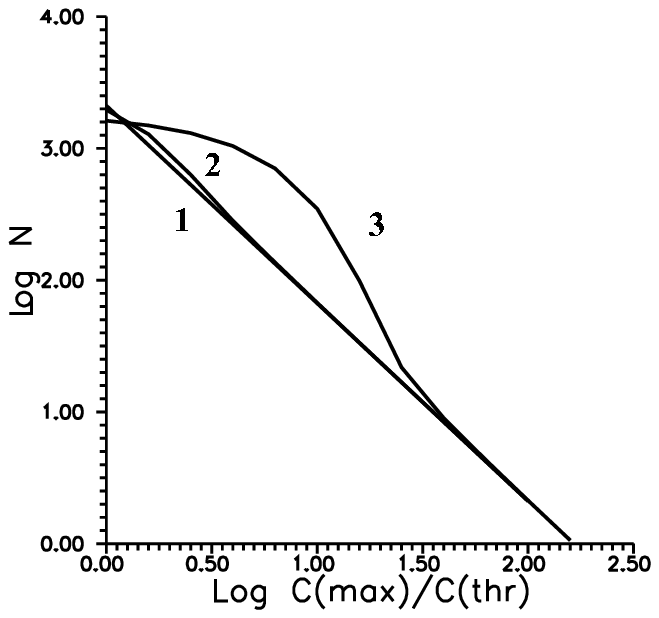}\,\epsfsize=05.5cm
\epsfbox{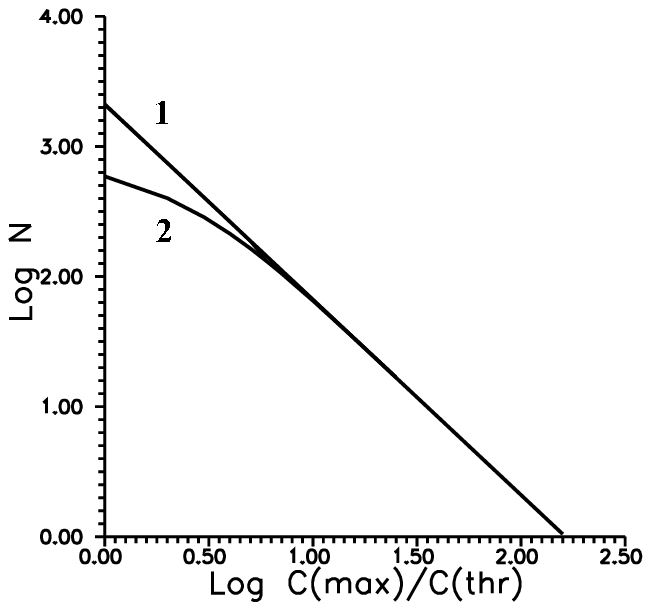}\
%,\epsfsize=04cm
%\vspace{12cm}
\caption[h]{{\bf a.} The curve $[\log N - \log C(max)/C(thr)]$ in presence of
stochastic errors, distributed according to normal
distribution with an average error $\Delta_1$ in units of a
threshold;
1 - straight line with slope 3/2, corresponding
to $\Delta_1=0$; 2 - curve with $\Delta_1=1$;
3 - curve with $\Delta_1=10$.
{\bf b.} Same as in {\bf a} for normal logarithmic distribution,
$\Delta$ determines a logarithm of number of thresholds, as an
average error;
 1 - straight line with a slope 3/2, corresponding
to $\Delta=0$; 2 - curve with $\Delta=1$; $C(max)$ is the peak intensity
of the burst; $C(thr)$ is a corresponding threshold value, from \cite{bk3}.
}
\label{grbfig6}
\end{figure}

The possibility of decisive role of selection effects
(incompleteness of data, statistical errors in estimation of
luminosity in presence of the threshold) are illustrated in figs.
\ref{grbfig5},\ref{grbfig6}. The incompleteness of data
influences the distribution of such a well studied stars as
solar type G stars, even larger effects are expected for such
short transients as GRB. The comparative dependence of the average
$V/V_{max}$
as a function of a cutoff for G stars and GRB is given in
fig.\ref{grbfig5} \cite{har95}. They show a striking qualitative similarity.

\noindent
The influence of statistical errors in presence of the threshold
was analyzed in \cite{bk3}. The  $\log N\, -\, \log S$ curves in
presence of statistical errors on the level of average 10
thresholds are represented in fig.\ref{grbfig6}a for the normal
distribution of the counts, and in fig.\ref{grbfig6}b for similar
distribution of their logarithms. The distribution in
fig.\ref{grbfig6}a has a similarity with the BATSE distribution in
fig.\ref{grbfig3}.

\section{Basement of a cosmological GRB origin:  optical
afterglows and red shifted lines}

The $X$-ray afterglows detected by Beppo-SAX gave a possibility of
optical identification and obtaining optical spectra. These
spectra have shown a strikingly large red shifts $z$, up to 4.5,
indicating to the cosmological origin of GRB and their enormous
energy outputs. In most cases the red shifts have been measured in
the host galaxies which are very faint. The list of red shift
measurements is given in the Table 2 from \cite{kul01}.
This table is completed by the trigger number and fluence
from 4B catalogue \cite{cat4}, and fluence for the GRB
from other references. The
spatial position of the optical afterglow sources in the host
galaxies can be found in \cite{kul00}.

Huge energy output during a short time (0.1 - few 100 seconds)
create problems for the cosmological interpretation. In some cases
like GRB 990123 the isotropic energy of the burst ($2.3\cdot
10^{54}$ ergs) exceeds the rest mass energy of the Sun. The
mechanisms discussed above are able to produce much less that one percent of
this amount.

\subsection{Collimation}

To avoid a huge energy production strong collimation is suggested
in the radiation of GRB. In the "cannon-ball" model \cite{ddd}
the bulk motion Lorentz factor is $\Gamma \approx 10^2\,-\,10^3$,
leading to collimation factor $\Omega \approx
10^{-4}\,-\,10^{-6}$. Analysis of GRB collimation have been done
in \cite{rho}. The main restriction to the collimation angle
follows from the analysis of the probability of appearance of the
orphan optical afterglow, which most probably have low or no
collimation. The absence of any variable orphan afterglow in a
search poses the following restrictions. It was expected to detect
$\sim 0.2$ afrerglows, if bursts are isotropic,
so the absence of orphan afterglows suggests
$\Omega_{opt}/\Omega_{\gamma}<<½ 100$,
which is enough to rule out the most extreme collimation scenarios.
At radio wavelengths published source counts and variability studies
have been used in \cite{pl98}
to place a limit on the collimation angle,
$\theta_{\gamma} \geq 5^{\circ}$.
Because radio afterglows last into the
non-relativistic phase of the GRB remnant
evolution, the radio afterglows are expected to radiate essentially
isotropically, and
the orphan afterglow limits on
radio $\Omega_r/\Omega_{\gamma}$ immediately imply a limit on
$\Omega_{\gamma}$ itself.

\begin{table}[h]
\caption{GRB Host Galaxies, Redshifts and Fluences (June 2001)}
\small{
\begin{center}
{\begin{tabular}{lllllll}
\hline\noalign{\smallskip}

 Trigger&
  GRB &
  $R$ mag &
  Redshift &
  Type $^a$&
  Fluence$^e$&
   Ref.\\
number&&&& & erg/cm$^2$&\\
\noalign{\smallskip}
\hline
\noalign{\smallskip}

& 970228     &   25.2 &  0.695   & e & 10$^{-5}$&\cite{hc97}  \\
6225& 970508     &   25.7 &  0.835   & a,e & $3.5\cdot 10^{-6}$(3+4)&  \\
6350& 970828     &   24.5 &  0.9579  & e & $7 \cdot 10^{-5}$&\cite{gg98} \\
6533& 971214     &   25.6 &  3.418   & e & $ 10^{-5}$(3+4)&   \\
6659& 980326     &   29.2 &$\sim$1?  &   & $6.3 \cdot 10^{-7}$(3+4)& \\
6665& 980329     &   27.7 &$<$3.9    & (b)&$7.1\cdot 10^{-5}$(3+4)& \\
6707& 980425 $^c$&   14   &  0.0085  & a,e&$4.4 \cdot 10^{-6}$&\cite{gv98}  \\
6764& 980519     &   26.2 &          &  &$9.4 \cdot 10^{-6}$(all 4) &    \\
& 980613     &   24.0 &  1.097   & e &$1.7\cdot 10^{-6}$&\cite{gn112} \\
6891& 980703     &   22.6 &  0.966   & a,e&$5.4\cdot 10^{-5}$(3+4)&\cite{gn126} \\
7281& 981226     &   24.8 &          & &$2.3\cdot 10^{-6}$(3+4)&    \\
7343& 990123     &   23.9 &  1.600   & a,e &$5.1 \cdot 10^{-4}$&\cite{gn224}\\
7457& 990308 $^d$&$>$28.5 &          &    &$1.9\cdot 10^{-5}(3+4)$ \\
7549& 990506     &   24.8 &  1.30    & e  &$2.2 \cdot 10^{-4}$&\cite{gn306}  \\
7560& 990510     &   28.5 &  1.619   & a &$2.6 \cdot 10^{-5}$&\cite{gn322} \\
& 990705     &   22.8 &  0.86    & x&$\sim 3 \cdot 10^{-5}$ &\cite{mas00}    \\
& 990712     &   21.8 &  0.4331  & a,e&&  \\
& 991208     &   24.4 &  0.7055  & e&$\sim 10^{-4}$ &\cite{gn450} \\
7906& 991216     &   24.85&  1.02    & a,x&$2.1\cdot 10^{-4}$(3+4)& \\
7975& 000131     &$>$25.7 &  4.50    & b &$\sim 10^{-5}$& \cite{gn529}  \\
& 000214     &        &0.37--0.47& x&$\sim 2 \cdot 10^{-5}$&\cite{gn557}    \\
& 000301C    &   28.0 &  2.0335  & a  & $\sim 4\cdot 10^{-6}$ &\cite{gn568}\\
& 000418     &   23.9 &  1.1185  & e&$1.3\cdot 10^{-5}$&\cite{gn642}  \\
& 000630     &   26.7 &          &   &$2\cdot 10^{-6}$&\cite{gn736}  \\
& 000911     &   25.0 &  1.0585  & e &$5 \cdot 10^{-6}$&\cite{gn791}  \\
& 000926     &   23.9 &  2.0369  & a &$2.2 \cdot 10^{-5}$&\cite{gn802}  \\
& 010222     &$>$24   &  1.477   & a& brightest of  &\cite{gn959} \\
&&&&&BeppoSAX&\\
\hline \\
\end{tabular}
}

\textsc{Notes}: \\
$^a$ e = line emission, a = absorption, b = continuum break, x = x-ray \\
$^c$ Association of this galaxy/SN/GRB is somewhat controversial \\
$^d$ Association of the OT with this GRB may be uncertain \\
$^e$The number of BATSE peak channel is indicated in brackets, from \cite{cat4}, \\
otherwise the estimation of bolometric fluence, and Reference are indicated\\
%}
\end{center}
}
%\label{Tab1a}
\end{table}

\subsection{Prompt optical afterglow in GRB 990123}

The light curve of the prompt optical afterglow looks similar to that of the
main GRB itself. It may be seen in the afterglow of
GRB 990123, which was catched by optical observations 22 seconds
after the onset of the burst \cite{ak1,ak2}.
GRB 990123 was detected by BATSE on 1999 January 23.407594
\cite{gn224}.
The event was strong and consisted of a multi-peaked
temporal structure lasting $\ge$100 s, with significant spectral
evolution.  The T50 and T90 durations are 29.82 ($\pm$ 0.10) s and 63.30
($\pm$ 0.26) s, respectively. So, optical emission of this GRB was detected
before its maximum in gamma region, and moreover, the gamma ray
maximum almost coincides with the optical one.
The data from \cite{ak1}, representing unfiltered CCD optical data are
given in the Table 3 for 23 January, the last 3 lines give R-
magnitude of the transient from \cite{kul99}.

\begin{table}
\caption{Observations of the optical transient of GRB 990123 January 23}
\begin{center}
\medskip
\begin{tabular}{ccccc}
\hline\noalign{\smallskip}
{time}&
{collection time}&
{magnitude}\\
\noalign{\smallskip}
\hline\\
    9:47:18.3 &   5 secs.&    11.82\\
    9:47:43.5 &   5 secs.&     8.95\\
    9:47:08.8 &   5 secs.&    10.08\\
    9:51:37.5 &  75 secs.&    13.22\\
    9:54:22.8 &  75 secs.&    14.00\\
    9:57:08.1 &  75 secs.&    14.53\\
13:37:20.3 &  &            18.70 $\pm$ 0.04\\
13:51:03.6 &  &        18.78 $\pm$ 0.04\\
14:02:56.5 &  &        18.75 $\pm$ 0.06\\
\hline\\
\end{tabular}
\end{center}
\end{table}
\noindent
That indicates to
the structure in which radiation comes almost simultaneously in
all energy bands, what is possible in the expanding transparent
plasma cloud illuminated by the gamma ray flux. Such model was
proposed for a GRB explosion near the neutron star surface in
\cite{bki89}, when the galactic origin of GRB was overwhelming.
It is quite unclear how to construct a similar model for
the cosmological GRB. The observed optical luminosity, related to
the red shift $z=1.61$ reaches $L_{opt} \approx 4 \cdot 10^{49}$
ergs/s, what is about 5 orders of magnitude brighter than optical
luminosity of any observed supernova.

\subsection{Optical afterglows from a re-radiation by the
interstellar gas}

\begin{figure}[h]
\epsfsize=06cm
{\centerline{\epsfbox{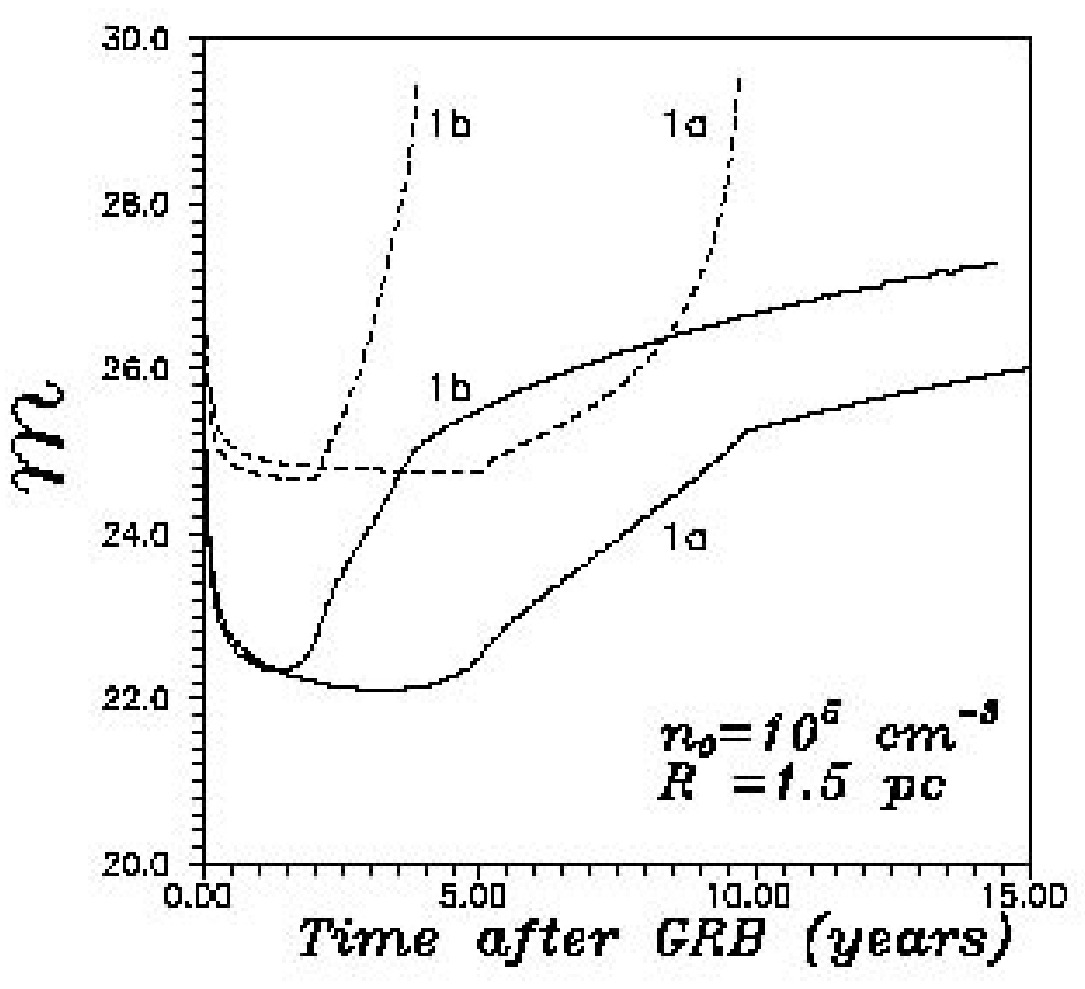}}}
%\vspace{12cm}
\caption[h]{
The magnitudes of the counterparts (upper limit - solid line,
 lower limit - dashed line) as a function of time after burst
for GRB with total flux near the Earth $F_{\rm GRB} = 10^{-4}$ erg
cm$^{-2}$:
1a. - for the case E = $10^{52}$ erg; $n_0= 10^5$ cm$^{-3}$;
1b -  for the case E = $10^{51}$ erg;  $n_0= 10^5$ cm$^{-3}$, from
\cite{bkt}}
\label{grbfig7}
\end{figure}

It was shown first in \cite{pa}, that properties of GRB afterglows
are explained better under suggestion that GRB source is situated
in a dusty star forming region with a high gas density.
Interaction of mighty GRB pulse with the surrounding gas with
a density $n=10^4\,-\,10^5$ cm$^{-3}$ create a specific form of the
optical afterglow, lasting up to few tens of years. The
calculations of light curve and spectrum of such afterglow
have been done in \cite{bkt}. Some results are represented in
fig. \ref{grbfig7}.
It is shown, that counterparts of cosmological GRB due to
interaction of gamma-radiation with dense interstellar media
are "long-living" objects, existing for years after GRB.
To distinguish GRB counterpart from a supernova event,
having similar energy output, it is necessary to take into
account its unusual light curve and spectrum. In the optical
region of the spectrum the strongest emission lines are
$H_{\alpha}$ and $H_{\beta}$.
Discovery of even one optical counterpart of
GRB with properties described above would give an opportunity
to probe the density of the interstellar medium
around the burst, and therefore would give an indication
of the burst progenitor.

\noindent

\section{Correlations}

Up to now no correlation had been found between GRB
distribution and large scale structure of the universe. This could
be connected with a insufficient angular resolution of GRB (few
degrees for most events). Combined with the error analysis on the BATSE catalog
it is concluded in \cite{sbg00} after using of the 4th (current) BATSE
catalog (2494 objects), that nearly $10^5$ GRBs will be needed to make a
positive detection of the two-point angular correlation function
at this angular scale, if the BATSE catalog is assumed to be a volume-limited sample
 up to $z \simeq 1$.

 Analysis of correlations between red shift, hardness and
luminosity have been done in \cite{sha}. In all 112 GRBs with both
luminosities and red shifts have been used, while most red shifts
were derived from the luminosities and the measured peak fluxes.

It was found that the hardness ratio between BATSE channels 3 and 1 do
change significantly with luminosity in that the luminous bursts are
harder than faint bursts. This correlation is in accordance with
the suggestion \cite{s01}, based on statistical analysis (see
Section 3), that there is
such a strong excess of luminosity in hard GRB that they are seen
from larger red shifts and
overcomes the tendency of the uniform Euclidian sample.

No significant correlation between hardness and red shift was found, as
might have been expected for cosmological shifting of the peak
energy.  However,
as the low luminosity events must be nearby and the high luminosity events
tend to be very distant, the effect from the previous paragraph will
approximately offset the cosmological shift resulting in the lack of any
apparent correlation \cite{sha}.

\begin{figure}[h]
\epsfsize=06cm
{\centerline{\epsfbox{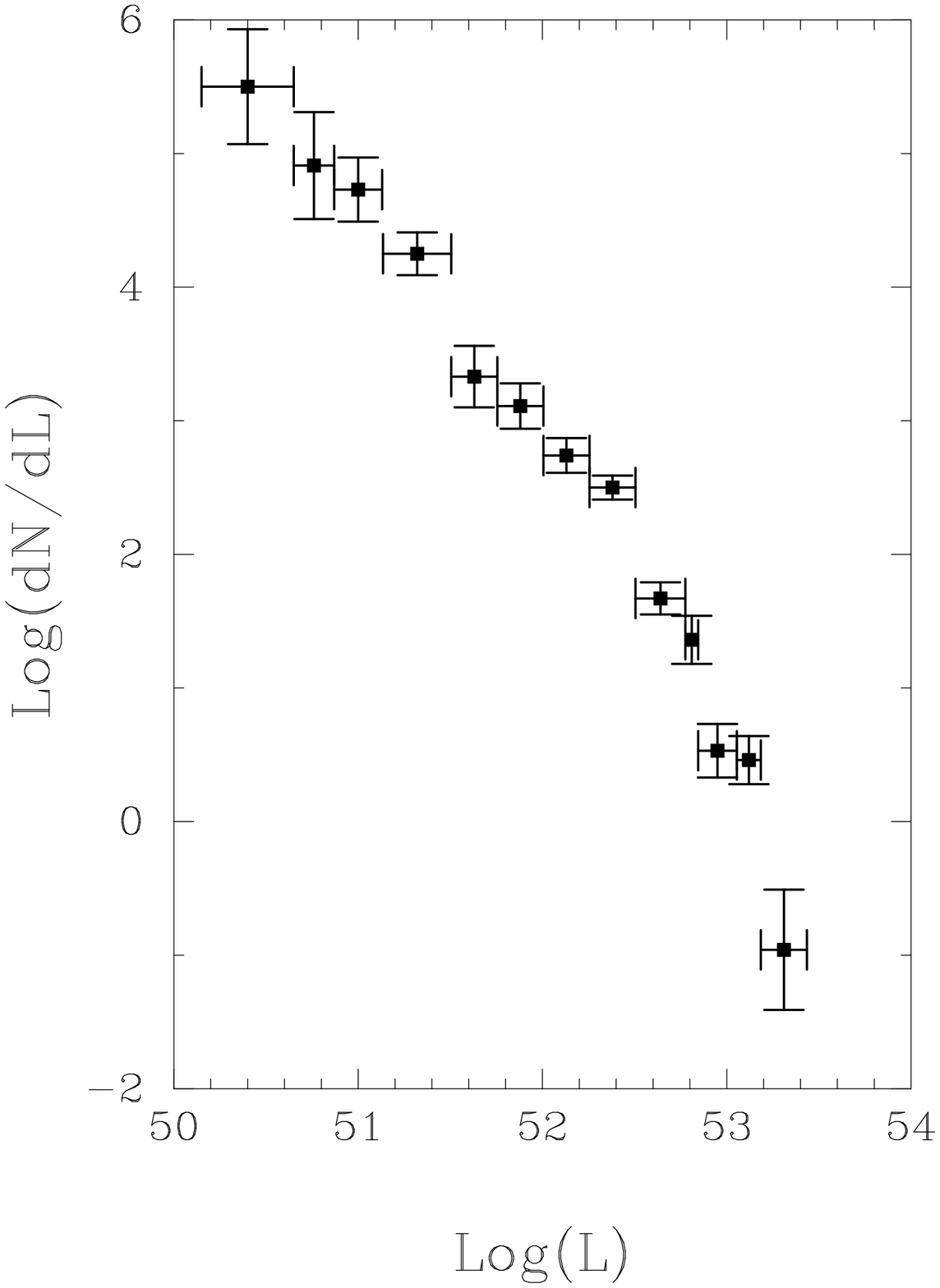}}}
\caption[h]{
The GRB luminosity function.  This measured luminosity function shows
the (arbitrarily scaled) number of bursts that appear within a luminosity
bin with width $10^{50}$ erg/s
as a function of luminosity.  The functional form is a broken power
law, from \cite{sha}.}
\label{grbfig8}
\end{figure}

Figure \ref{grbfig8} displays the
luminosity function, taken as the number of
bursts appearing within luminosity bins of width $10^{50}$ erg/s.
The luminosity
function appears as a broken power law with the break at $2 \times
10^{52}$ erg/s. The
dependence above the break is fitted to be scaling as $L^{-2.8 \pm 0.2}$,
while it scales as $L^{-1.7 \pm 0.1}$ below the break \cite{sha}.

Comparison of the red shifts and fluences from Table 2 shows no
correlation between distance and observed flux. Even in view of
large scattering of GRB power it looks rather unusual, and needs
considerable skill to explain this property in the cosmological
model.

\noindent

 \section{High-energy afterglow}

\begin{figure}[h]
\epsfsize=014cm
{\centerline{\epsfbox{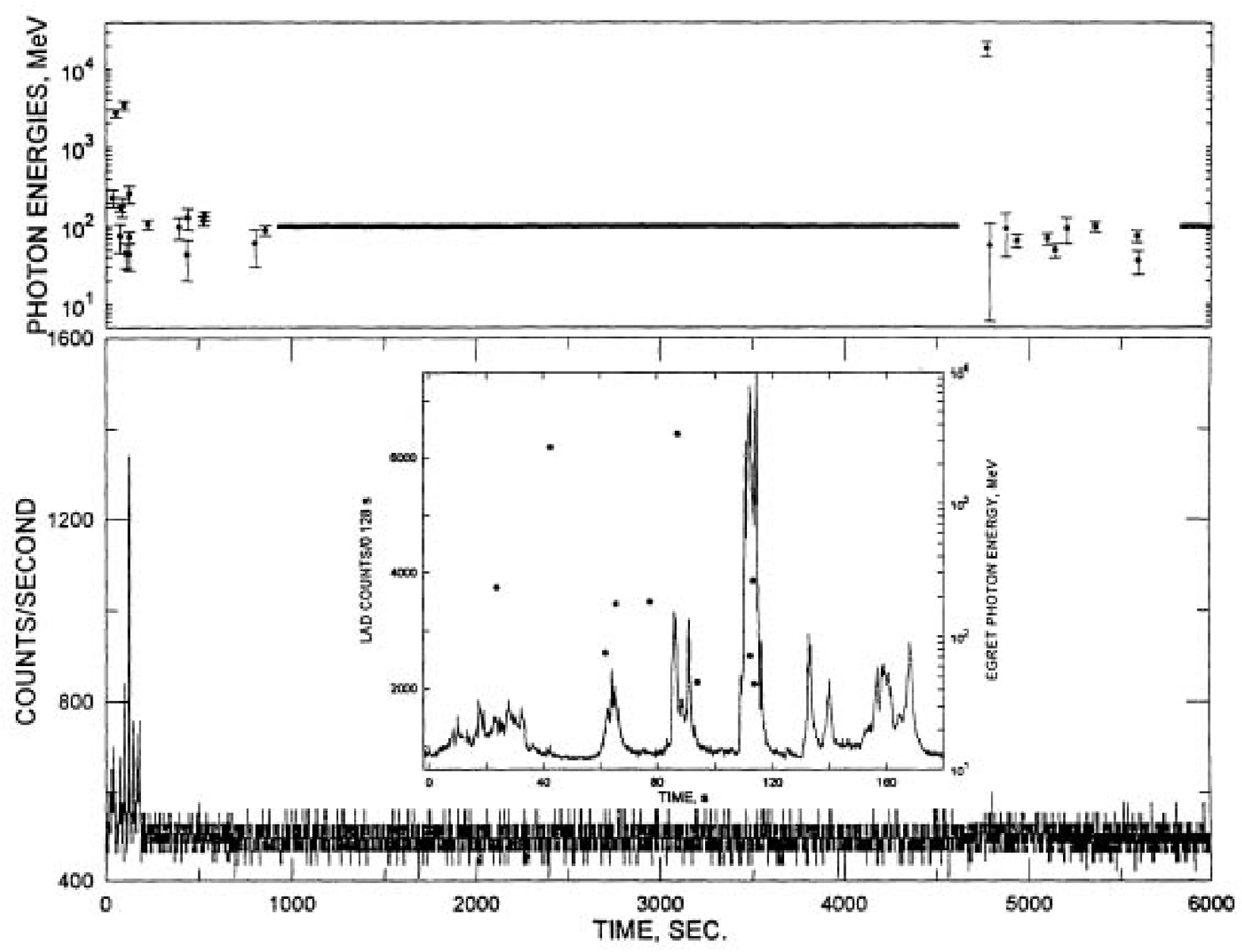}}}
\caption[h]{
The burst of 17 February 1994, observed to emit GeV photons up to 1.5 hours
after the initial outburst, as observed by EGRET experiment. The composite
figure includes data from EGRET, Ulysses and BATSE experiments, from \cite{fm95}.}
\label{grbfig9}
\end{figure}

EGRET observations on CGRO have shown that GRB emit also very hard
gamma photons up to 20 GeV \cite{fm95}. The number of GRB with
detected hard gamma radiation is about 10, from them 5 bursts with
registered photon energies over 100 MeV are given in Table 4 from
\cite{schneid}. Hard gamma emission, as a rule, continues longer
than the main (soft) gamma ray burst, up to 1.5 hours in the
GRB940217, see fig.\ref{grbfig9}.
Comparison of the angular aperture of EGRET and BATSE
leads to conclusion that hard gamma radiation could be observed in
large fraction (about one half) of all GRB. Spectral slope in hard
gamma region varies between (-2) and (-3.7), see Table 4, and
varies rapidly, becoming softer with time (GRB920622 in
\cite{schneid2}). Data about spectra of hard gamma radiation of
radio pulsars in Crab nebula and PSR B1055-52 \cite{crab,gpsr}
show similar numbers and variety. With account of non-pulsed
Crab spectrum the slope varies between (-1.78) and (-2.75),
what is close to GRB spectral slope.

\begin{table}
\caption{EGRET Energetic Gamma-Ray Bursts Observations, from
\cite{schneid}}
\begin{center}
\medskip
\begin{tabular}{ccccc}
\hline\noalign{\smallskip}
{Burst ID}&
{Max. Energy}&
{Duration}&
{Spectral}&
{Delayed}\\
{}&
{(GeV)}&
{Emission}&
{Function}&
{Emission}\\
\noalign{\smallskip}
\hline\\
 GRB910503 & $10$ & $84$ s & $E^{-2.2}$ & X \\
 GRB910601 & $0.314$ & $200$ s & $E^{-3.7}$ & X \\
 GRB930131 & $1.2$ & $100$ s & $E^{-2.0}$ & X \\
 GRB940217 & $18$ & $1.5$ h & $E^{-2.6}$ & X \\
 GRB940301 & $0.16$ & $30$ s & $E^{-2.5}$ &  \\
\hline\\
\end{tabular}
\end{center}
\end{table}

\noindent

 \section{Hard X-ray lines}

\begin{figure}
\epsfsize=012cm
\epsfbox{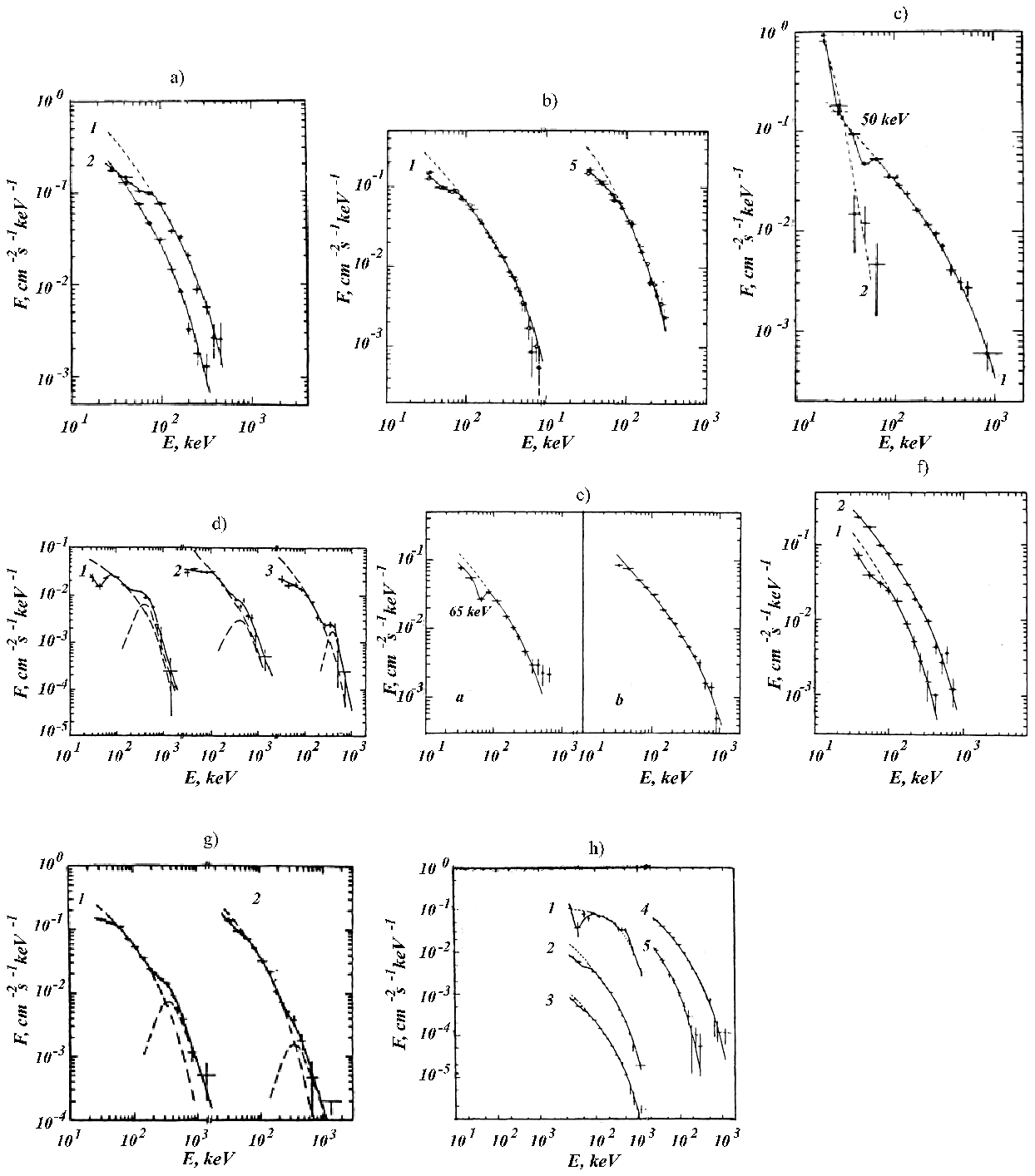}
%{\centerline{\epsfbox{grbfig1.eps}}}
%\vspace{12cm}
\caption[h]{KONUS data on the spectral variability of GRB:
$t_{i+1}>t_i$.
{\bf a.} 780914;
{\bf b.} 790325;
{\bf c.} 790329;
{\bf d.} 790402;
{\bf e.} 791101, $t_b>t_a$;
{\bf f.} 800105;
{\bf g.} 820825;
{\bf h.} 820827C, from
\cite{kon82}.}
\label{spectr}
\end{figure}

\begin{figure}
\epsfsize=06cm
\epsfbox{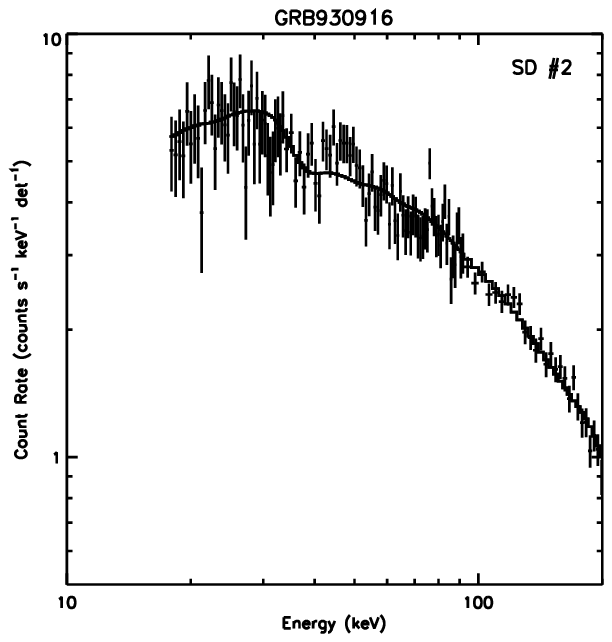}\,\epsfxsize=6.0cm
\epsfbox{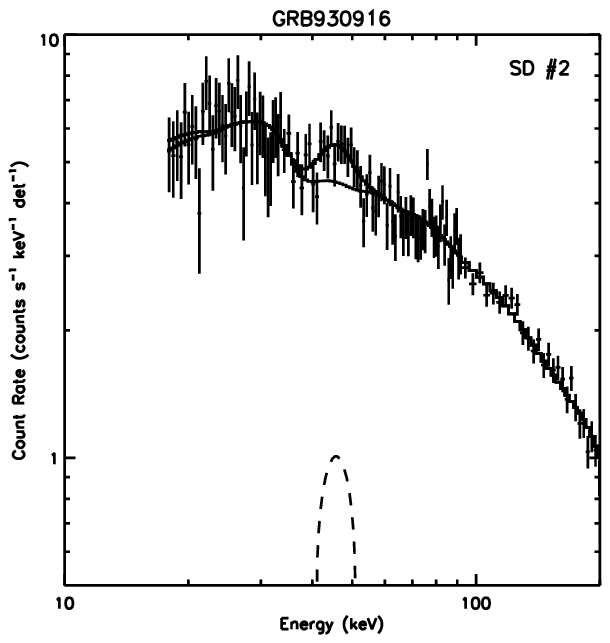}\,\epsfxsize=6.0cm
%{\centerline{\epsfbox{grbfig1.eps}}}
%\vspace{12cm}
\caption[h]{\small Data from the interval 22.144 to 83.200~s after the BATSE trigger
of GRB930916.
The plot shows the count rate data (points) and count rate models (histograms).
The count rate models are obtained by folding the photon models through
a model of the detector response.
The `bump' at 30 keV is expected from the K-edge of the iodine in NaI.
Left panel: best continuum-only fit to the data of SD~2.
The data show a clear excess above the model from 41 to 51 keV.
Right panel: A narrow spectral feature is added to the model: an emission line
at 45 keV improves $\chi^2$ by 23.1.   The width of the feature is due to the
detector resolution.   The solid histogram depicts the total count model; the
dashed histograms show the continuum and line portions separately,
from \cite{briggs}.}
\label{briggs1}
\end{figure}

\begin{figure}
\epsfsize=06cm
\epsfbox{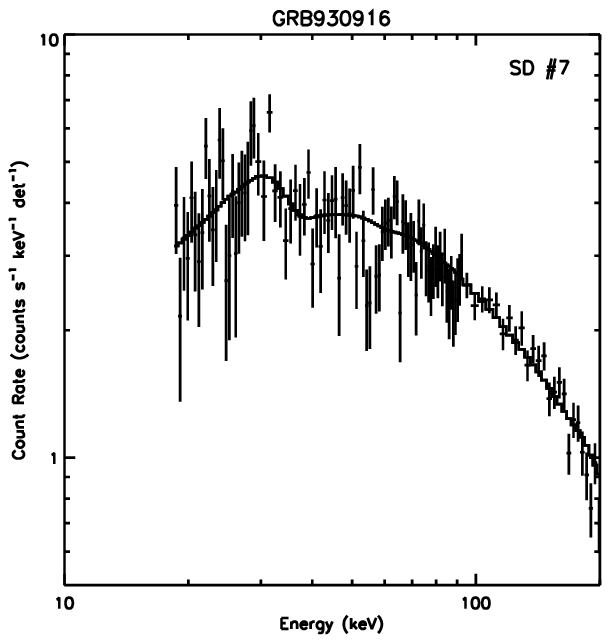}\,\epsfxsize=6.0cm
\epsfbox{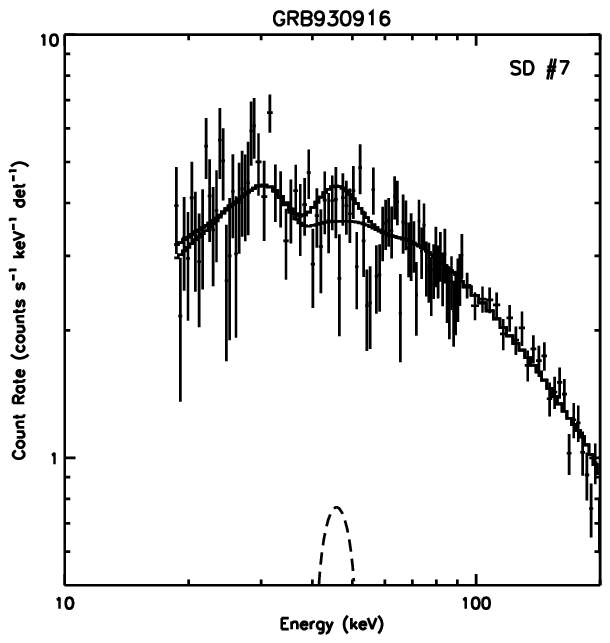}\,\epsfxsize=6.0cm
%{\centerline{\epsfbox{grbfig1.eps}}}
%\vspace{12cm}
\caption[h]{Left panel: best continuum-only fit to the data of SD~7 for
the same time interval of GRB930916 as used in fig. \ref{briggs1}.
Adding a line results in no $\chi^2$ improvement.
Right panel:
The continuum model is still a fit, but a line at the strength indicated
by the data of SD~2 is imposed.
The model in the region of the putative line is clearly above the data
and $\chi^2$ is increased (compared to the continuum-only fit of
the left panel) by
9.7, rather than decreased, from \cite{briggs}.}
\label{briggs2}
\end{figure}

\begin{figure}[h]
\epsfsize=08cm
{\centerline{\epsfbox{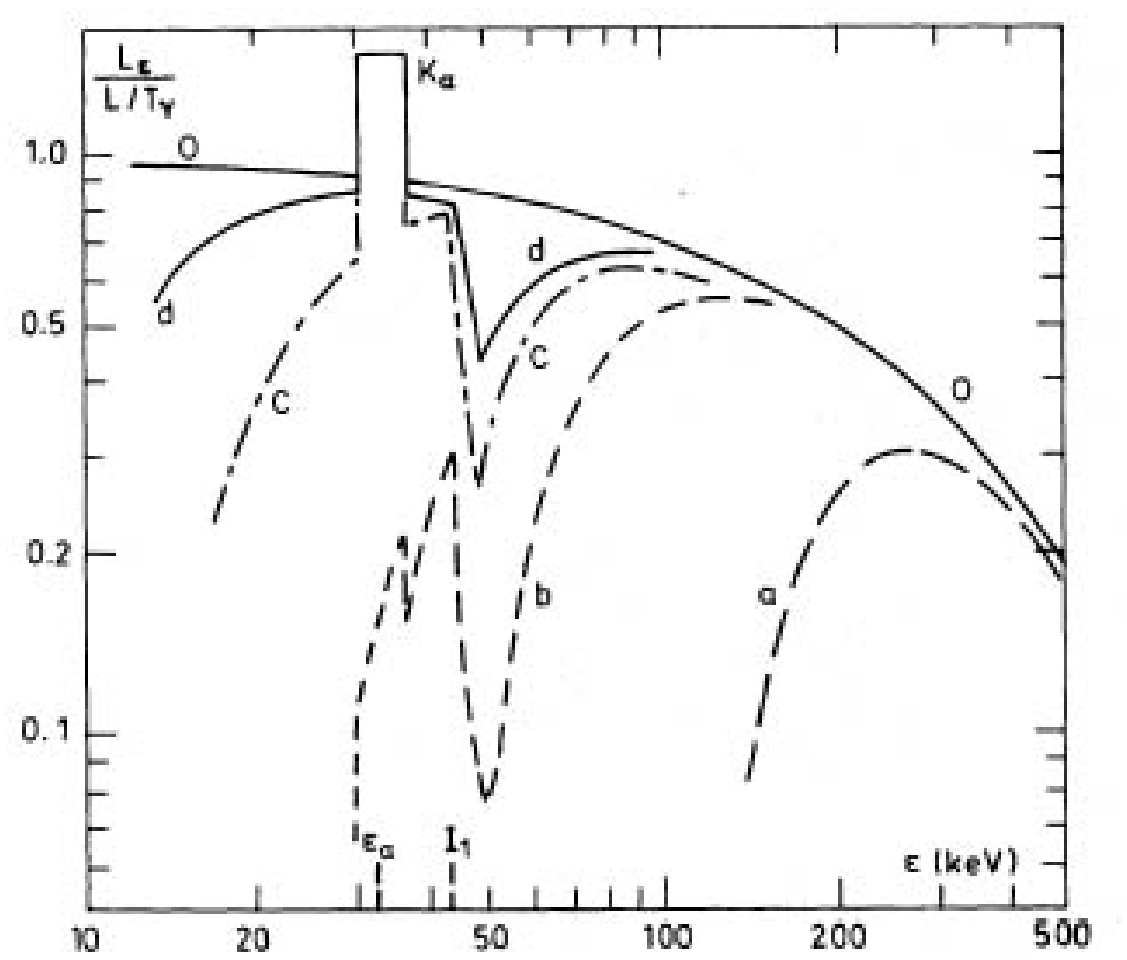}}}
\caption[h]{The time evolution of the absorption line formed by a
fraction of barium ions in the iron plasma with $X_{Ba}=1/300$
in the spectrum of GRB; $t_a<t_b<t_c<t_d$; "0" is the spectrum of
radiation falling on on the cloud, from \cite{bki89}.
}
\label{bki}
\end{figure}

Hard gamma-ray lines in GRB spectra have been discovered by KONUS
group \cite{kon82}. They had been interpreted there as cyclotron
lines, and have been seen in 20-30\% of the GRB. These spectra had
shown a distinct variability: the visible absorption decreases with time,
what may be seen in fig.~\ref{spectr}.
BATSE detectors had lower spectral resolution, and for a some time
no spectral features have been found in their data. Later
publications appeared where possibility of existence of hard X-ray
spectral features in GRB spectra was found (see \cite{briggs} and references
therein).
In \cite{briggs} 13 statistically significant line candidates have been found from
117 GRBs.
One of the best cases for detecting a line is GRB~941017 in which
the data from two detectors are consistent. In some GRBs the line
was found only by one detector, while it was not statistically
significant in the other one. The example of this situation for
GRB930916 is
given in figs. \ref{briggs1},\ref{briggs2}. The conclusion of this
paper is, that until we have a better understanding of these
apparent inconsistencies between the
data collected from different detectors,
the reality of all of the BATSE line candidates is unclear.
Note that spectra in figs.\ref{briggs1},\ref{briggs2} have been
obtained 20 s after the trigger, and according to \cite{kon82} the
lines are the strongest at the beginning of the burst.
The interpretation of hard spectral feature in the cosmological
model \cite{grblines} was based on the blue-shifted
($\Gamma=25-100$) spectrum of the gas cloud illuminated by the
gamma radiation of the fireball. Similar model was suggested in
\cite{bki89} for explanation of the lines observed by KONUS in the
model of an explosion near the surface of the neutron star and
formation of expanding cloud with $v/c=0.1-0.3$, see
fig.\ref{bki}, which explained the time evolution of the line
strength in fig.\ref{spectr}.


 \section{Discussion}

The cosmological origin of GRB create many problems for
construction of its physically realistic model. The main
difficulty is to combine huge energy production during short
time and in a small volume. The collimation should help in this
situation, but its angle cannot be too small. The investigation of
orphan optical bursts by all-sky optical monitoring could be very
useful for putting better limits for the collimation. It is very
important to obtain prompt optical spectra of the GRB afterglows
when the optical counterpart is still luminous, and to investigate
the polarization of the optical and X-ray afterglow for
clarification of the radiation mechanism.

\begin{figure}
\epsfsize=06cm
\epsfbox{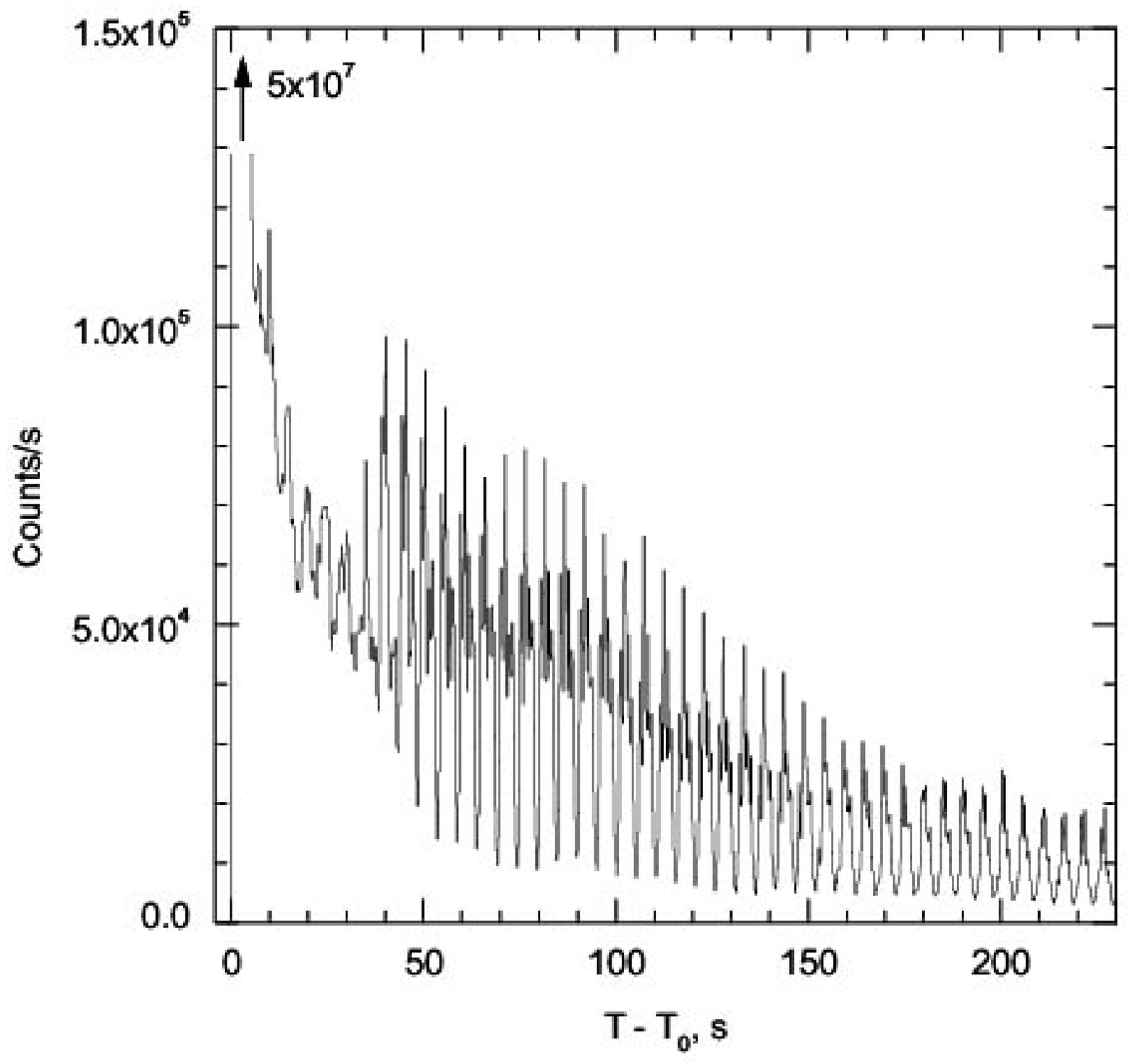}\,\epsfxsize=6cm
\epsfbox{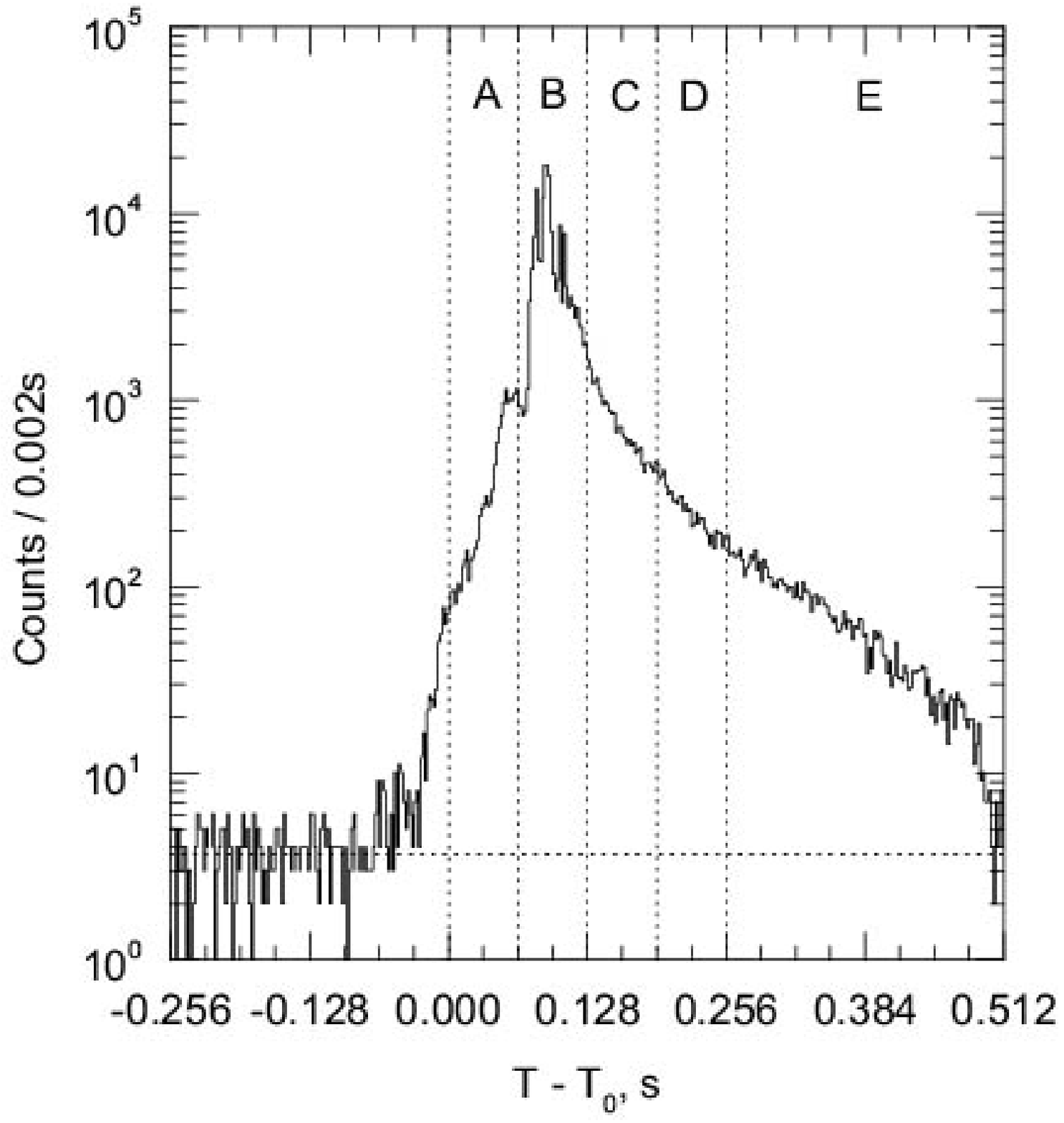}\,\epsfxsize=6cm
\caption[h]{{\bf a}. The giant 1998 August 27 outburst
of the soft gamma repeater SGR 1900 + 14.
 Intensity of the $E > 15$ keV radiation.
{\bf b}. The time history of the giant burst from the soft gamma repeater SGR
1627-41. on June 18, 6153 s UT corrected for
dead time. Photon energy $E > 15$ keV. The rise time is about 100 ms,
from \cite{sgr1,sgr2}}
\label{sgrm}
\end{figure}

It may happen, that GRB are not a uniform sample of objects, but
include phenomena of different origin. The statistical analysis
reveals at least two separate samples consisting of long ($>\sim
1$~s) and short bursts. Note, that optical afterglows and redshift
measurements have been done only for long bursts. Therefore, it is
not excluded that short bursts have different (may be galactic)
origin. It is interesting to compare the properties of short GRB
with giant bursts from soft gamma-repeaters (SGR), which are
situated inside the Galaxy. From the larger distance, when a usual
SGR activity is not visible, only giant bursts would be
registered, which without doubts could be attributed to the short
GRB. Two giant bursts are given in fig. \ref{sgrm} according to
KONUS-WIND observations \cite{sgr1,sgr2}. The existence of the giant
bursts in the SGR (3 in 4 firmly known SGR in the Galaxy and LMC)
implies a possibility for observation of these giant bursts, which
appear as short GRB, in other neighboring galaxies. The estimation
gives more than 10 expected "short GRB" of this type from M 31 and
other close neighbors \cite{bkvul99}.
The absence of any GRB projecting on the
local group galaxies may indicate that SGR are more close and less
luminous objects, than it is now accepted  \cite{bkvul99}.

\bigskip

\noindent
{\Large{\bf Acknowledgements}}
\smallskip

This work was partially supported by RFBR grant 02-02-16900 and
INTAS-ESA grant 120.
I am very grateful to Franco Giovannelli and other organizers of the
Workshop for support and kind hospitality, and to Olga Toropina
for help in preparation of this contribution.

\end{document}